\documentclass[12pt,ams]{article}%
\usepackage[nohead]{geometry}
\usepackage{geometry}
\usepackage{dsfont}
\usepackage{amsmath}
\usepackage{amsfonts}
\usepackage{amssymb}
\usepackage{graphicx}%
\usepackage{subfig}
\usepackage{float}
\usepackage{tikz}
\usepackage{epsfig}
\usepackage{hyperref}
\begin{document}

\date{}
\title{\textbf{Effective field theories for superconducting systems with multiple Fermi surfaces}}

\author{
\textbf{P. R. Braga.}$^{a}$\thanks{pedro.rangel.braga@gmail.com}\,\,,
\textbf{D. R. Granado}$^{a,c}$\thanks{diegorochagrana@uerj.br,\ Diego.RochaGranado@ugent.be}\,\,,
\textbf{M. S. Guimaraes}$^{a}$\thanks{msguimaraes@uerj.br}\,\,,
\textbf{C. Wotzasek}$^{b}$\thanks{clovis@if.ufrj.br}\,\,,
% \thanks{AGRADECIMENTOS}
  \\ \small \textnormal{$^{a}$ \it Departamento de F\'{\i }sica Te\'{o}rica, Instituto de F\'{\i }sica, UERJ - Universidade do Estado do Rio de Janeiro} \
   \\ \small \textnormal{\phantom{$^{a}$} \it Rua S\~{a}o Francisco Xavier 524, 20550-013 Maracan\~{a}, Rio de Janeiro, Brasil.}\
 \\ \small \textnormal{$^{b}$\it Instituto de F\'\i sica, Universidade Federal do Rio de Janeiro, 21941-972, Rio de Janeiro, Brazil.}\
 \\ \small \textnormal{$^{c}$ \it  Department of Physics and Astronomy,  Ghent University} \
\\ \small \textnormal{\phantom{$^{c}$} \it  Krijgslaan 281-S9, 9000 Gent, Belgium}}

\maketitle

%----------------------------------------------------------------------------------------
%	ABSTRACT
%----------------------------------------------------------------------------------------
 
\begin{abstract}

In this work we investigate the description of superconducting systems with multiple Fermi surfaces. For the case of one Fermi surface we re-obtain the result that the superconductor is more precisely described as a topological state of matter. Studying the case of more than one Fermi surface, we obtain the effective theory describing a time reversal symmetric topological superconductor. These results are obtained by employing a general procedure to construct effective low energy actions describing states of electromagnetic systems interacting with charges and defects. The procedure consists in taking into account the proliferation or dilution of these charges and defects and its consequences for the low energy description of the electromagnetic response of the system. We find that the main ingredient entering the low energy characterization of the system with more the one Fermi surface is a non-conservation of the canonical supercurrent triggered by particular vortex configurations.

\end{abstract}

%----------------------------------------------------------------------------------------
%	Se??o 1
%----------------------------------------------------------------------------------------
\section{Introduction}

In this work we investigate the low energy description of electromagnetic systems as a function of configurations of charged sources and defects. We specifically address the case when such systems are superconducting and may display multiple Fermi surfaces.

This present work was mainly motivated by the illuminating study developed  in \cite{Hansson:2004wca}, where the authors centered the analysis of a superconductor state on the proper identification of its low energy degrees of freedom. This lead to the conclusion that the effective low energy theory of a superconductor, usually described by Ginzburg-Landau theory, is more appropriately described by a topological $BF$ theory encoding the topological interactions of vortices and charges in the system. This strategy led us to consider the use of a procedure, known as the Julia-Toulouse approach (JTA) \cite{Grigorio:2009pi}, which naturally accommodates such a  focus on the behavior of the degrees of freedom. The JTA has the aim of constructing an effective field theory for gauge fields from semi-classical considerations about the collective dynamics of charged particles and defects in the system. The procedure is throughly reviewed in  \cite{Grigorio:2011pi} and has already been employed for the study of superconducting systems in relation to confinement  \cite{Guimaraes:2012tx}, see also \cite{Grigorio:2012jt, Guimaraes:2012ma} for related work.

Another motivation for the present work comes from the recent studies of Qi, Witten and Zhang \cite{Qi:2012cs}. These authors proposed an effective theory for a time reversal invariant topological superconductor in three spatial dimensions, $(3+1)D$ (for a review of topological materials see \cite{Qi:2011zya}).  Their construction involves consideration of  a time reversal topological insulator in $(4+1)D$ sandwiched between two boundaries, where each boundary sustain a $(3+1)D$ $s$-wave superconductor. This construction leads, upon dimensional reduction, to an effective description of topological superconductor characterized by a topological term describing a coupling between the electromagnetic field and the superconducting phase fluctuation. Such a coupling is mathematically the same as the one between an Abelian gauge field and an axion. One of the main results presented in their work is the realization of the phenomenon of anomaly inflow \cite{Callan:1984sa} due to the contribution of so called ``chiral vortices''.

In this work we provide a construction of the effective field theory of superconductor states that encompasses the observations in \cite{Hansson:2004wca}, thus recovering their findings. Further, we show that this construction can tackle the case of multiple Fermi surfaces as well, and we show that this naturally leads to the results of \cite{Qi:2012cs}. This construction thus highlight the fact that the main element characterizing the class of topological superconductor discussed in \cite{Qi:2012cs} is the presence of more than one Fermi surface, such that the single Fermi surface superconductor discussed in \cite{Hansson:2004wca} can be viewed as a special case. Furthermore we clarify the role played by the different vortices in the system providing a precise relation between them and the anomaly of the supercurrent.  We observe that an apparent paradox in the results of \cite{Qi:2012cs} was reported in \cite{Stone:2016pof}, where the authors pointed out that Majorana fermions, being uncharged, could not account for the anomaly inflow and proceed to propose a solution to the anomaly imbalance (for a review about the role of Majorana fermions in topological superconductor see \cite{Leijnse:2012tn, Beenakker:2011np}). In the present work we add to these findings by showing that the non-conservation of the supercurrent is a consequence of the existence of a particular configuration of vortices that do not carry electromagnetic flux. Nevertheless this is just an apparent anomaly restricted to the non-conservation of the canonical supercurrent, since gauge symmetry is maintained throughout the whole procedure and the total electric current in the system is conserved, as it should.

This work is organized as follows. In section 2 we present the general description a superconductor state from the point of view of the JTA relying on a parametric condensation or dilution of vortices in the system. This section will provide the main concepts and ingredients for the following sections. In section 2.1 we present our first main result which is the recovering of the results of \cite{Hansson:2004wca} under our formalism. In section 3 we discuss the case of a system displaying two Fermi surfaces. A superconductor is characterized by the dynamics of excitations near a Fermi surface. The key observation here is to allow for charge transfer between Fermi surfaces. This gives rise to a non-trivial topological interaction between different Fermi surfaces and results to be the origin of the topological properties obtained in the effective theory put forward by \cite{Qi:2012cs}. In section 4 we generalize the results for an arbitrary number of Fermi surfaces. In section 5 we present our conclusions.

\section{Effective theory for a superconductor and condensation}

 \noindent Consider the Euclidean action defining a gauge field interacting with a classical source  in $4D$
\begin{align}
\label{maxwell}
S_{em} = \int d^4x \left( \frac{1}{4} F_{\mu\nu}F_{\mu\nu} - iqA_{\mu}J_{\mu}  \right)
\end{align}
where $F_{\mu\nu} = \partial_{\mu}A_{\nu} - \partial_{\nu}A_{\mu}$ and $J_{\mu}$ is a classical current. This action describes the electromagnetic field interacting with an external source with electric charge $q$. We define the partition function of the system, which is the generating functional of gauge field correlation functions,  by coupling the system with an external auxiliary current $j_{\mu}$, carrying charge $e$, and integrating over gauge field configurations as well as summing over the classical configurations of the external sources $J_{\mu}$
 \begin{align}
 \label{pathint-maxwell}
 Z[j] = \sum_{J}  \int {\cal D} A \delta[\partial_{\mu}J^{\mu}] e^{-\int d^4x \left( \frac{1}{4} F_{\mu\nu}F_{\mu\nu} - iqA_{\mu}J_{\mu}  \right)} e^{-ie\int d^4x \; A_{\mu}j_{\mu}  }
 \end{align}
where the integral measure for the gauge field must be defined with an appropriate gauge fixing, which we omit in the notation. We also inserted a delta function constraining the classical source to be conserved in order do maintain gauge invariance. This delta function may be exponentiated with the help of an auxiliary field $\theta$, resulting
\begin{align}
 \label{pathint-maxwell-2}
 Z[j] = \sum_{J}  \int {\cal D} A {\cal D} \theta e^{-\int d^4x \left( \frac{1}{4} F_{\mu\nu}F_{\mu\nu} - iq \left( A_{\mu} + \frac{1}{q} \partial_{\mu}\theta \right) J_{\mu}\right)} e^{-ie\int d^4x \; A_{\mu}j_{\mu}  }
 \end{align}
In this form, there is no constraint in the current $J_{\mu}$ and gauge symmetry is realized as
\begin{align}
 \label{gaugesymmetry}
A_{\mu} &\rightarrow  A_{\mu}  + \partial_{\mu} \chi \nonumber\\
\theta &\rightarrow \theta - q\chi
 \end{align}
 The sum over the current $J_{\mu}$ configurations is so far unspecified. We can give a prescription to perform this sum and in doing so we will define different behaviors for the resulting gauge field correlation functions. In that sense, prescribing the ensemble of current configurations define different gauge theories. In fact, as example, consider the extreme case of $J_{\mu} =0$ as the only  configuration. This is just free Maxwell theory.  Another more interesting case is to consider $J_{\mu}$ as a continuous field in such a way that the sum is simply replaced by a path integral, i.e., $\sum_{J} \rightarrow \int {\cal D} J$. This is tantamount to consider the system as a perfect electric condensate. The result is that $J_{\mu}$ turns into a Lagrange multiplier forcing the gauge field to vanish, which is just the Meissner effect in a perfect superconductor (with zero penetration lenght).
 
 \noindent We can study this process of dilution and condensation under an equivalent perspective. It is known that the process of charge condensation can be viewed from a dual point of view, where defects (vortices in this superconductor example) in the system goes through a process of dilution. In order to see this, we make use of the Poisson summation formula
 \begin{align}
  \label{poissonformula}
  \sum_{J} \delta[J^{\mu}(x) - \eta^{\mu}(x)] = \sum_{K} e^{i2\pi \int d^4 x \eta^{\mu}(x) K_{\mu}(x) }
  \end{align}
  Where $J$ and $K$ are $1$-currents and $\eta$ is a $1$-form. The generalization for $p$-currents and $p$-forms in $D$ dimensions is straightforward. This formula can be proved by going to the lattice formulation and starting with the original Poisson summation formula, with $J$ and $K$ integer valued variables and $\eta$ a real valued function \cite{Grigorio:2009pi}.  Note that the extreme examples of complete dilution and complete condensation are consistently represented in this formula. If $J_{\mu} =0$ is the only configuration on the left hand side we get a $\delta[\eta^{\mu}(x)] $, which is the result if we make $\sum_{K} \rightarrow \int {\cal D} K$ in the right hand side. Contrariwise, if we make $\sum_{J} \rightarrow \int {\cal D} J$ on the left hand side, we get $1$, which is the result if $K_{\mu} =0$ is the only configuration in the right hand side. Thus, we conclude that a complete dilution of $J_{\mu} $ corresponds to a complete condensation of $K_{\mu}$ and vice-versa. This suggests that $K_{\mu}$ stands for the defects, dual to the charges $J_{\mu} $.

\noindent These extreme examples of complete condensation or complete dilution can be seen as the end points of a physical process of condensation of the electric charges and illustrates the point of view we want to convey here. Mathematically, we can consider a diluted electric charge configuration as an ensemble of $1$-currents, that is, distribution-valued $1$-forms. For instance, for a single electric charge we have 
 \begin{align}
 \label{linhamundo}
 J^{\mu}(x) = \int d\tau \frac{dy^{\mu}}{d\tau} \delta^4 (x-y(\tau))
 \end{align}
 and the summation $\sum_{J} $ amounts to an integral over all the possible charge's worldlines $y(\tau)$ weighted by the charge's action $S(y(\tau))$ describing the charge's dynamics. For many point charges we would have a sum over the worldlines of all the charges. For a continuous distribution of charges we would have a continuous source, whose sum over different ensemble configurations is defined by a path integral weighted by an action $S(J)$\footnote{Treating $J$ as an external current leads to the interpretation of the action $S(J)$ as generating contact terms for the n-point functions of the fundamental fields \cite{Closset:2012vp}} . The condensation, whose extreme cases were described above, is thus an operation that maps an ensemble of $1$-currents into an ensemble of $1$-forms. This operation specifies a physical process that connects different theories, describing the system in different phases.

\noindent In order to see this, let's rewrite the partition function \eqref{pathint-maxwell-2}. We define the sum over currents making explicit the weight $e^{-S_J}$ defining the ensemble of currents. Also, we insert in the path integral the unit 
 \begin{align}
 \label{unitpoisson}
 \int {\cal D }\eta\; \delta[J^{\mu}(x) - \eta^{\mu}(x)] = 1
 \end{align}
 obtaining
\begin{align}
\label{genfunc}
  Z[j]  = \sum_{J}  \int {\cal D} A {\cal D} \theta {\cal D} \eta\delta[J^{\mu}(x) - \eta^{\mu}(x)]  e^{-\int d^4x \left( \frac{1}{4} F_{\mu\nu}F_{\mu\nu} - iq \left( A_{\mu} + \frac{1}{q} \partial_{\mu}\theta \right) \eta_{\mu}\right)} e^{-ie\int d^4x \; A_{\mu}j_{\mu}  }e^{-S_{\eta}} 
\end{align}
where, due to delta function, we replaced $J\rightarrow \eta$ everywhere in the argument. Now,  using the Poisson summation formula (\ref{poissonformula}), we obtain the equivalent formulation
\begin{align}
\label{genfunc2}
Z[j]  = \sum_{K}  \int {\cal D} A {\cal D} \theta {\cal D} \eta e^{-\int d^4x \left( \frac{1}{4} F_{\mu\nu}F_{\mu\nu} - iq \left( A_{\mu} + \frac{1}{q} \partial_{\mu}\theta + \frac{2\pi}{q} K_{\mu}\right) \eta_{\mu}\right)} e^{-ie\int d^4x \; A_{\mu}j_{\mu}  }e^{-S_{\eta}} 
\end{align}

\noindent We can now have a more clear understanding of the meaning of the ensemble action $S_{\eta}$. In order to perform the integral over $\eta$ we have to specify a form for this action. Consider first the simplest choice of putting $S_{\eta} = 0$. In this case, $\eta$ is a Lagrange multiplier that imposes 
\begin{align}
\label{etalag}
 A_{\mu} = -\frac{1}{q} \partial_{\mu}\theta - \frac{2\pi}{q} K_{\mu} 
\end{align}
since $K$ is a $1$-current, it defines localized lines in space. This equation just says that the gauge field $A_{\mu}$ is restricted to flux filaments and thus we see that $K$ stands for vortices in the system. If $K$ dilutes the vortices disappears and the system becomes a perfect superconductor with zero penetration length for the magnetic field. On the other hand, if $K$ condenses then we recover free Maxwell theory with the field $K$ as the gauge field. 

\noindent Turning on the action $S_{\eta}$ adds more structure to this setting, but the physics is the same. Consider the derivative expansion
\begin{align}
 \label{Seta}
  S_{\eta} = \frac{1}{2M^2} \eta_{\mu}\eta^{\mu} + \frac{1}{2M_1^4} \eta_{\mu}\partial^2 \eta^{\mu} + \cdots
\end{align}
Where $M$ and $M_1$ are mass parameters. This expansion is justified if we want to describe the low energy fluctuations of the field $\eta$. Considering only the first term in this action, which corresponds to choosing the lowest order term in this derivative expansion, we can readily solve the path integral for $\eta$, using its equation of motion (since $\eta$ is an auxiliary field in this approximation):
\begin{align}
\label{eqmoteta}
\eta_{\mu} =  iq M^2 \left( A_{\mu} + \frac{1}{q} \partial_{\mu}\theta + \frac{2\pi}{q} K_{\mu} \right)
\end{align}
 and we obtain 
 \begin{align}
 \label{dilutecharges3}
 Z[j] = \sum_{K}  \int {\cal D} A \int {\cal D} \theta\;  e^{-\int d^4x \left( \frac{1}{4} F_{\mu\nu}F_{\mu\nu} + \frac{q^2M^2}{2}\left(  A_{\mu} + \frac{1}{q} \partial_{\mu}\theta + \frac{2\pi}{q} K_{\mu}\right)^2 \right)} e^{-ie\int d^4x \; A_{\mu}j_{\mu}  } \end{align}
which is the action for the electromagnetic response in a superconductor with penetration length $\sim 1/M$. 

\subsection{The topological theory of a superconductor} 

\noindent The previous description of the superconductor state is suitable to be cast in a manifestly topological field theory description. The superconductor is usually regarded as a state resulting of a spontaneously broken gauge theory, but the very concept of gauge symmetry breaking is misleading to say the least, since gauge symmetry is not a symmetry, but a redundancy in the theoretical description. The observation that the superconductor is a topological state of matter was made in \cite{Hansson:2004wca} and in this section we cast this conclusion under the light of the discussion of the previous section.

 \noindent An important clue for the interpretation of the superconductor state as a topological state of matter is the identification of its low energy degrees of freedom. There are three types of excitations that can be read from the partition function \eqref{dilutecharges3}: quasiparticles with coupling $e$, vortices and massive vector particles (massive photons). As pointed out in \cite{Hansson:2004wca}, even if naively one may consider the quasiparticles to carry a classical charge $e$, this is not so because these quasiparticles are charged with respect to a massive vector field, so that the classical charge, measured by the assimptotic flux of the massive photon, dies away for distances $\gg 1/M$. Therefore the quasiparticles with coupling $e$ are classically neutral. 

 \noindent But these quasiparticles do carry a topological or quantum charge coming from an Aharonov-Bohn interaction with the vortices. The energy density remains finite if asymptotically the field strength $F_{\mu\nu}$ and the mass term $\left(A_{\mu} + \frac{1}{q} \partial_{\mu}\theta + \frac{2\pi}{q} K_{\mu}\right)^2$  vanishes. It follows that at large distances we have \footnote{Note that in the present formulation $\theta$ is a regular field and not an angular variable}
\begin{align}
 \label{vortexflux}
 \oint_C dx^{\mu} A_{\mu} = \frac{2\pi}{q} \oint_C dx^{\mu} K_{\mu} = \frac{2\pi n}{q} 
 \end{align}
 Where we used the fact that $\oint_C dx^{\mu} K_{\mu} = n \in \mathds{N}$ is the linking number between line $C$ and the vortex line defined by the $1 $-current $K$. Therefore, the vortex carry a flux $\Phi= \frac{2\pi n}{q} $ and we identify $\Phi_0 = \frac{2\pi}{q}$ as the fundamental vortex flux. If we make a quasiparticle with coupling $e$ go around a vortex carrying fundamental flux $\Phi_0$, it will gain an Aharonov-Bohn phase   
 \begin{align}
  \label{AB}
  \theta_{AB} = e\oint_c dx^{\mu} A_{\mu} =e\Phi_0 =  \frac{2\pi e}{q} 
  \end{align}
For a superconductor $q=2e$ is the charge of a Cooper pair and in this case there is a nontrivial  phase
\begin{align}
  \label{AB2}
  \theta_{AB} = \pi
  \end{align}
As thoroughly discussed in \cite{Hansson:2004wca} this is the fundamental property that characterizes the superconductor as a topological state. The effective theory describing the superconductor state is a topological field theory describing the intersection of the world lines of the quasiparticles and vortices. This is given by the $BF$ theory defined by the action \cite{Blau:1991, Schwarz:1979} 
\begin{align}
  \label{BF}
S_{SC} = \int d^4 x \left(\frac{1}{\pi} \varepsilon^{\mu\nu\rho\sigma}b_{\mu\nu}\partial_{\rho}a_{\sigma} - a_{\mu}j^{\mu} - b_{\mu\nu}j^{\mu\nu}_V\right)
\end{align}
where $j^{\mu\nu}_V = \varepsilon^{\mu\nu\rho\sigma}\partial_{\rho}a_{\sigma}$ is the vortex current and $j^{\mu} = \varepsilon^{\mu\nu\rho\sigma}\partial_{\nu}b_{\rho\sigma}$ is the quasiparticle current. Topological field theories have become very important in the description of topological materials, see for instance\cite{Qi:2008ew, Hansson:2011rc, Wang:2010xh}

%%%%%%%%%%%%%%%%%%%%%%%%%%% 
% \subsection{Derivation of the BF theory}
%%%%%%%%%%%%%%%%%%%%%%%%%%%

\noindent Here we show how the $BF$ action \eqref{BF} can be derived by following the steps of the first section. Coupling the gauge field with a current with electric charge of a Cooper pair $q=2e$, the action in the partition function \eqref{genfunc2} reads:
\begin{equation}
S=\int d^4x\left(\frac{1}{4}F_{\mu\nu}F_{\mu\nu}-2ieA_\mu\eta_\mu-i\theta\partial_\mu\eta_\mu-2\pi iK_\mu\eta_\mu+ieA_\mu j_\mu\right)+S_\eta
\label{18}
\end{equation}
An important point to notice is that the vortices in this theory are defined by the vortex current
\begin{equation}
\label{vc}
J_V^{\mu\nu}=\epsilon^{\mu\nu\rho\sigma}\partial_\rho K_\sigma
\end{equation}
We can implement this information in the theory by the identity
\begin{equation}
\label{vortex-field}
\sum_Kf(K)=\sum_{J_V}\int\mathcal{D}a_{GF}\delta(J_V^{\mu\nu}-\epsilon^{\mu\nu\rho\sigma}\partial_\rho a_\sigma)f(a)
\end{equation}
where $f(x)$ is an arbitrary function. For a proof of this statement, see appendix \ref{proof}. The field $a$ in this formula is understood to be gauge fixed by $\partial_\mu a^\mu=0$. This expression effectively replaces $K$ by $a$ in all expressions. From this we obtain:
\begin{equation}
Z(j)=\sum_{J_V}\int\mathcal{D}A\mathcal{D}\eta\mathcal{D}\theta\mathcal{D}a \mathcal{D}b e^{iS}
\end{equation}
with
\begin{equation}
S=\int d^4x\left(\frac{1}{4}F_{\mu\nu}F_{\mu\nu}-2ei\eta_\mu(A_\mu+\frac{1}{2e}\partial_\mu\theta+\frac{\pi}{e}a_\mu)+ieA_\mu j_\mu
+ i(b_{\mu\nu}J_V^{\mu\nu}-b_{\mu\nu}\epsilon^{\mu\nu\rho\sigma}\partial_\rho a_\sigma)\right)+S_\eta
\end{equation}
where $b$ is the Lagrange multiplier implementing the condition \eqref{vortex-field}.

Now the superconductor is parametrized by the vortex current $J_V$ such that: if $J_V=0$, the Lagrange multiplier $b$ will force $\epsilon\partial a=0$. Therefore, $a=0$, since $\partial a=0$ (or $a=\partial\phi$ if the gauge is not fixed) and we obtain a pure superconductor state upon integration over $\eta$. On the other hand, if $J_V$ condenses ($\sum_J\rightarrow\int\mathcal{D}J$) it becomes a Lagrange multiplier forcing $b=0$, as a result $a$ also becomes a Lagrange multiplier forcing $\eta=0$ and we recover the usual Maxwell action coupled to $j_\mu$. This is the non-superconducting state.

Returning to the intermediate cases, where there are vortices in the system but they are not condensed, we can perform the integration over $\eta$. We consider the case where  a saddle point approximation is justified, so that we just have to solve the equations of motion:
\begin{equation}
\frac{\delta S}{\delta\eta_\mu}=-2ei(A^\mu+\frac{1}{2e}\partial^\mu\theta+\frac{\pi}{e}a^\mu)+\frac{\delta S_\eta}{\delta\eta_\mu} = 0
\end{equation}
for the case:
\begin{equation}
S_\eta=\int d^4x \frac{1}{2M^2}\eta_\mu\eta^\mu
\end{equation}
It follows
\begin{equation}
\eta_\mu=2M^2ei(A_\mu+\frac{1}{2e}\partial_\mu\theta+\frac{\pi}{e}a_\mu)
\end{equation}
replacing this solution in the action we have:
\begin{eqnarray}
S_\eta- 2ei\int d^4x\eta^\mu(A_\mu+\frac{1}{2e}\partial_\mu\theta+\frac{\pi}{e}a_\mu)
= 2e^2M^2\int d^4x(A_\mu+\frac{1}{2e}\partial_\mu\theta+\frac{\pi}{e}a_\mu)^2
\end{eqnarray}
It is convenient to define
\begin{eqnarray}
B_\mu&\equiv&A_\mu+\frac{1}{2e}\partial_\mu\theta+\frac{\pi}{e}a_\mu\\
F_{\mu\nu}(B)&=&F_{\mu\nu}(A)+\frac{\pi}{e}F_{\mu\nu}(a)
\end{eqnarray}
Then the action becomes:
\begin{align}
S=\int d^4x  &\left[ \frac{1}{4}F^2_{\mu\nu}(B)+\frac{\pi^2}{4e^2}F^2_{\mu\nu}(a)-\frac{2\pi}{e}B_\nu\partial_\mu F^{\mu\nu}(a)+2e^2M^2B^2\right.\nonumber\\
&+\left.ieB_\mu j^\mu-i\pi a_\mu j^\mu+ib_{\mu\nu}J_v^{\mu\nu}-ib_{\mu\nu}\epsilon^{\mu\nu\rho\sigma}\partial_\rho a_\sigma\right]
\end{align}
integrating $B$ we will get only higher order contribution in $\partial F(a)$ for the effective action. Therefore, at low energies the action reads:
\begin{equation}
S_{eff}=\int d^4x\left(-i\pi a_\mu j^\mu +ib_{\mu\nu}J_v^{\mu\nu}-ib_{\mu\nu}\epsilon^{\mu\nu\rho\sigma}\partial_\rho a_\sigma\right)
\end{equation}
or redefining $a\to\frac{1}{\pi}a$ and $b\to-b$ we get
\begin{equation}
\label{BFtopological}
S_{eff}=i\int d^4x\left(\frac{1}{\pi}b_{\mu\nu}\epsilon^{\mu\nu\rho\sigma}\partial_\rho a_\sigma-a_\mu j^\mu-b_{\mu\nu}J_V^{\mu\nu}\right)
\end{equation}
which we recognize as the topological $BF$ theory defined by the action \eqref{BF}.

%%%%%%%%%%%%%%%%%%%%%%%%%%%%%%%%%%%%%%%%%%%%%
\section{The case of two Fermi surfaces: effective field theory for a topological superconductor}
%%%%%%%%%%%%%%%%%%%%%%%%%%%%%%%%%%%%%%%%%%%%%
In order to extend these ideas to the description of a superconductor in a system with more than one Fermi surface, we start with the same partition function as before 
 \begin{align}
 \label{pathint-maxwell-3}
 Z[j] = \sum_{J}  \int {\cal D} A \delta[\partial_{\mu}J^{\mu}] e^{-\int d^4x \left( \frac{1}{4} F_{\mu\nu}F_{\mu\nu} - iqA_{\mu}J_{\mu}  \right)} e^{-ie\int d^4x \; A_{\mu}j_{\mu}  }
 \end{align}
but we note that the current can be written as the sum of the contributions coming from each Fermi surface. For the case of two Fermi surfaces, we have 
\begin{align}
\label{currents-2Fermi}
J^{\mu} = \frac{1}{2}\left(  J_1^{\mu}+ J_2^{\mu} \right)
\end{align}
The important point to observe is that only the full current needs to be conserved. Physically, there may be an electric flow between the different Fermi surfaces. We may represent this locally by instantons contributions 
\begin{align}
  \label{currents-2Fermi2}
 \partial_{\mu}  J_1^{\mu} &= \rho \nonumber\\
 \partial_{\mu}  J_2^{\mu} &= -\rho 
\end{align} 
where $\rho$ is the instanton contribution leading to the non-conservation of  the charge localized in each Fermi surface. One can picture the instantons as ``holes'' connecting the Fermi surfaces, from where the electric charge may leak.  We can rewrite the partition function in order to display this information
\begin{align}
 \label{pathint-maxwell-4}
 Z[j] = \sum_{J_1, J_2, \rho}  \int {\cal D} A \delta[\partial_{\mu}J_1^{\mu}-\rho]\delta[\partial_{\mu}J_2^{\mu}+\rho] e^{-\int d^4x \left( \frac{1}{4} F_{\mu\nu}F_{\mu\nu} - i\frac{1}{2}qA^{\mu}\left( J_1^{\mu}+ J_2^{\mu} \right) \right)} e^{-ie\int d^4x \; A_{\mu}j_{\mu}  }
 \end{align} 
 Note the sum over instantons ensemble configurations. One way to understand the necessity of this sum is to consider the limit in which $\rho$ condenses; in this case the Fermi surfaces effectively superpose, since there is free electric current flow between them. Mathematically, the sum over $\rho$ becomes an integral and we are left with the only constraint of total charge conservation. The opposite limit of instantons dilution is also instructive. In this case we have independent conservation of charges in each Fermi surface, since there are no ``holes''. 
 
 % % % % % % % % % % % %

We can recast this formulation in terms of the total current $J_{\mu}$, \eqref{currents-2Fermi}, and the relative current 
\begin{align}
  \label{currents-rel}
 \bar{ J}^{\mu} = \frac{1}{2}\left(  J_1^{\mu} - J_2^{\mu} \right)
\end{align} 
The partition function \eqref{pathint-maxwell-4} becomes
\begin{align}
 \label{pathint-maxwell-4a} 
 Z[j] &= \sum_{J, \bar{J}, \rho}  \int {\cal D} A \delta[\partial_{\mu}J^{\mu}]\delta[\partial_{\mu}\bar{J}^{\mu}-\rho] e^{-\int d^4x \left( \frac{1}{4} F_{\mu\nu}F_{\mu\nu} - iqA^{\mu}J^{\mu} \right)} e^{-ie\int d^4x \; A_{\mu}j_{\mu}  }\nonumber\\
 &= \sum_{J, \bar{J}, \rho}  \int {\cal D} A  {\cal D} \theta  {\cal D} \bar{\theta} \; e^{-\int d^4x \left( \frac{1}{4} F_{\mu\nu}F_{\mu\nu} - iq\left( A_{\mu} + \frac{1}{q} \partial_{\mu} \theta \right)J_{\mu} - i\bar{J}_{\mu}  \partial_{\mu}\bar{\theta} - i\rho\bar{\theta}\right)} e^{-ie\int d^4x \; A_{\mu}j_{\mu}  }
 \end{align} 
Where in the second line we have exponentiated each constraint in the delta functions with fields $\theta$ and $\bar{\theta}$.  We note that, since only the total current couples with the gauge field,the system displays the following charge symmetry
\begin{align}
  \label{currents-redef}
  J_{\mu} &\rightarrow  -J_{\mu}; \;\;\;
  \bar{J}_{\mu} \rightarrow  \bar{J}_{\mu} ;\;\;\;
  A_{\mu} \rightarrow - A_{\mu} \nonumber\\
  \theta &\rightarrow - \theta ; \;\;\;
  \bar{\theta} \rightarrow  \bar{\theta} ; \;\;\;
  \rho \rightarrow \rho
\end{align} 
Now, following the same procedure that led us to eq.\eqref{genfunc2}, we insert units, one for each current,  in the form of eq.\eqref{unitpoisson}, introducing the corresponding fields $\eta$ and $\bar{\eta}$. The Poisson identity eq.\eqref{poissonformula} will replace the currents $J$ and $\bar{J}$ by their Poisson dual $K$ and $\bar{K}$. We then obtain 
\begin{align}
 \label{pathint-maxwell-5}
 Z[j] = \sum_{K, \bar{K}, \rho} \int {\cal D} A  {\cal D} \theta {\cal D} \eta  {\cal D} \bar{\theta} {\cal D} \bar{\eta} \; e^{- S_{K,\bar{K}}  }e^{-ie\int d^4x \; A_{\mu}j_{\mu}  }e^{-S_{\eta, \bar{\eta}}} 
 \end{align} 
where
\begin{align}
\label{action2fermi}
S_{K,\bar{K}} = \int d^4x \left( \frac{1}{4} F_{\mu\nu}F_{\mu\nu} - iq \left( A_{\mu} + \frac{1}{q} \partial_{\mu}\theta + \frac{2\pi}{q} K_{\mu}\right) \eta_{\mu} - i \left( \partial_{\mu}\bar{\theta} + 2\pi \bar{K}_{\mu}\right)\bar{\eta}_{\mu}  -i\rho \bar{\theta}\right)
\end{align} 
With the use of eq. \eqref{vortex-field} applied to $K$ and $\bar{K}$ this formulation can be equivalently rewritten in terms of two vortex currents, $J_V$ and $\bar{J}_V$, in the form 
\begin{align}
 \label{pathint-maxwell-6}
 Z[j] = \sum_{J_V, \bar{J}_V, \rho} \int {\cal D} A  {\cal D} \theta {\cal D} \eta  {\cal D} \bar{\theta} {\cal D} \bar{\eta} {\cal D} a {\cal D} b  {\cal D} \bar{a} {\cal D} \bar{b}\; e^{- S  }e^{-ie\int d^4x \; A_{\mu}j_{\mu}  }e^{-S_{\eta, \bar{\eta}}} 
 \end{align}  
\begin{align}
\label{action2fermi-2}
S &= \int d^4x \left( \frac{1}{4} F_{\mu\nu}F_{\mu\nu} - iq \left( A_{\mu} + \frac{1}{q} \partial_{\mu}\theta + \frac{2\pi}{q} a_{\mu}\right) \eta_{\mu} - i \left( \partial_{\mu}\bar{\theta} + 2\pi \bar{a}_{\mu}\right)\bar{\eta}_{\mu}  -i\rho \bar{\theta} \right. \nonumber\\ 
&\left. + ib_{\mu\nu}J_V^{\mu\nu} - ib_{\mu\nu}\epsilon^{\mu\nu\rho\sigma}\partial_{\rho}a_{\sigma} +i\bar{b}_{\mu\nu}\bar{J}_V^{\mu\nu} - i\bar{b}_{\mu\nu}\epsilon^{\mu\nu\rho\sigma}\partial_{\rho}\bar{a}_{\sigma}  \right)
\end{align} 
As before, in order to proceed, we have to specify a form for the action $S_{\eta, \bar{\eta}}$, so we will write it in the most general form as a derivative expansion, respecting the symmetries (note that $\eta$ and $\bar{\eta}$ have the same symmetries as $J$ and $\bar{J}$, respectively)
\begin{align}
   \label{etaaction}
S_{\eta,\bar{\eta}}=\frac{1}{2M^2}\eta_{\mu}\eta_{\mu}+\frac{1}{2m^2}\bar{\eta}_{\mu}\bar{\eta}_{\mu}+\frac{1}{\Lambda^{6}}\bar{\eta}_{\mu}\epsilon^{\mu\nu\rho\sigma}\eta_{\nu}\partial_{\rho}\eta_{\sigma}
\end{align}
Where $M$, $m$ and $\Lambda$ are mass parameters. 

At this point we can comment on the physical meaning of this formulation. The system defined by \eqref{pathint-maxwell-6}, \eqref{action2fermi-2} and \eqref{etaaction} can be understood as a function of the configurations of the vortex currents $J_V$ and $\bar{J}_V$. The system will display different states depending on the condensation or dilution of these currents. This is more clearly seen in the extreme situations of complete dilution or condensation. We thus recognize four main situations
\begin{itemize}
\item \emph{$J_V$ and  $\bar{J}_V$ condense}, that is, become fields to be integrated over in the path integral. In this case these currents act as Lagrange multipliers forcing $b$ and $\bar{b}$ to be zero. This in turn renders $a$ and  $\bar{a}$ into Lagrange multipliers that forces $\eta$ and $\bar{\eta}$ to zero, thus recovering the free Maxwell theory. This is the expected result of condensation of vortices leading to a destruction of the superconducting state.

\item\emph{ $J_V$ condenses and  $\bar{J}_V$ dilutes}. The dilution corresponds to consider only the configurations of the system in which $\bar{J}_V = 0$. In this case also the superconducting state is destroyed due to the condensation of $J_V$. The dilution of $\bar{J}_V $ implies that $\bar{a} = 0$ and the resulting theory for $\bar\theta$ decouples from the system.

\item \emph{$J_V$ dilutes and  $\bar{J}_V$ condenses}. As $J_V \rightarrow 0$ the system becomes superconducting, since $a = 0$ and the integration over $\eta$ provides a finite penetration length for the photon. The condensation of $\bar{J}_V$ implies $\bar{\eta} = 0$ and the corresponding sector has no effect in the electromagnetic dynamics.  This is therefore the state corresponding to a normal superconductor, as described in the last section.

\item \emph{$J_V$ and  $\bar{J}_V$ dilute}. When both currents dilute, we have a full topological superconductor. The dilution of $J_V$ is responsible for the establishment of the superconducting phase and the dilution of $\bar{J}_V$ provides the conditions for its topological properties. This case will be discussed below.

\end{itemize}

Employing a saddle point approximation, solving the equations of motion of \eqref{etaaction} for $\eta$ and $\bar\eta$ and taking only the lowest orders terms we obtain
\begin{eqnarray}
\eta_{\mu}&\approx& iqM^2B_{\mu}\\
\bar{\eta}_{\mu}&\approx& im^2\bar{B}_{\mu},
\end{eqnarray}
where the fields $B_{\mu}$ and $\bar{B}_{\mu}$ are defined as
\begin{align}
\label{bi}
B_{\mu} &= A_{\mu} + \frac{1}{q} \partial_{\mu}\theta + \frac{2\pi}{q} a_{\mu}.\nonumber\\
\bar{B}_{\mu} &= \partial_{\mu}\bar{\theta} + 2\pi \bar{a}_{\mu}.
\end{align}
Replacing this in the action \eqref{etaaction}, we finally have 
\begin{eqnarray}
\label{actionTSC2fermi}
S &= \int d^{4}x\; \left( \frac{1}{4}F_{\mu\nu}(A)F^{\mu\nu}(A) + \frac{q^{2}M^{2}}{2}B^{2}_{\mu} + \frac{m^{2}}{2}\bar{B}^{2}_{\mu}  -iq^{2}\frac{M^4m^{2}}{\Lambda^6}\bar{B}_{\mu}\epsilon^{\mu\nu\rho\sigma}B_{\nu}\partial_{\rho}B_{\sigma} - i\rho\bar{\theta}\right. \nonumber\\ 
&\left. + ib_{\mu\nu}J_V^{\mu\nu} - ib_{\mu\nu}\epsilon^{\mu\nu\rho\sigma}\partial_{\rho}a_{\sigma} +i\bar{b}_{\mu\nu}\bar{J}_V^{\mu\nu} - i\bar{b}_{\mu\nu}\epsilon^{\mu\nu\rho\sigma}\partial_{\rho}\bar{a}_{\sigma}  \right).
\end{eqnarray}

We can add one more piece of  information in the construction of the effective action by noting that the sum over instantons configurations $\rho$ can be performed, following Polyakov \cite{Polyakov:1975rs, Polyakov:1976fu, Polyakov:1996nc} considering a ``dilute gas of instantons'' approximation (see also \cite{Guimaraes:2012ma}). A generic instanton configuration will contribute to the partition function  with a factor $e^{-\frac{I}{e^2}} $, where $I$ represents the value of the action in the instanton configuration. This term will represent the probability of creation of an instanton. More precisely, the number of instantons inside a volume $V$ is 
\begin{align}
\label{instanton-contribution}
V \mu e^{-\frac{I}{e^2}},
\end{align}
where $\mu$  has dimension of inverse of volume, and $V\mu$ is a region of the material where an instanton can be located. If an instanton is located at spacetime point $x_{0}$, we have $\rho=\pm \delta^4(x-x_{0})$, with $+$ or $-$ sign for instantons and anti-instantons, repectively. The partition function for the instantons can be evaluated as
\begin{align}
\label{polyakov}
\sum_{instanton} e^{ \int d^4x\; i\rho(x)\bar{\theta}(x)} 
&=\sum^{\infty}_{n_{+}=0}\frac{ \left(\mu e^{\frac{-I}{ e^{2} } }\right)^{n_{+}} }{n_{+}!} \prod_{i=1}^{n_{+}} \int d^4x_{0i}\; e^{i\bar{\theta}(x_{0}) }\times\nonumber\\
&\times\sum^{\infty}_{n_{-}=0}\frac{ \left(\mu e^{\frac{-I}{ e^{2} } }\right)^{n_{-}} }{n_{-}!} \prod_{j=1}^{n_{-}} \int d^4x_{0j}\; e^{-i\bar{\theta}(x_{0})}\nonumber\\
&= e^{\int d^4x \; \tilde{\rho}\;cos(\bar{\theta}(x))}.
\end{align}
Where
\begin{align}
\label{expressiaon-of-rho}
\tilde{\rho}=2\mu e^{\frac{-I}{e^{2}}}.
\end{align}
This gives a very enlightening interpretation for the Josephson term in the effective action of a topological superconductor. It is indeed the contribution describing the flow of current tunneling between the Fermi surfaces characterized by the phase difference $\bar{\theta}$.

Therefore, the full effective action describing the topological superconductor in the case of two Fermi surfaces has the form 
\begin{eqnarray}
\label{topSC2FS-b}
S_j &=& S +ie\int d^4x \; A_{\mu}j_{\mu}  \nonumber\\
&=& \int d^{4}x\; \left( \frac{1}{4}F_{\mu\nu}(A)F^{\mu\nu}(A) + \frac{q^{2}M^{2}}{2}\left( A_{\mu} + \frac{1}{q} \partial_{\mu}\theta + \frac{2\pi}{q} a_{\mu}\right)^{2} + \frac{m^{2}}{2}\left(\partial_{\mu}\bar{\theta} + 2\pi \bar{a}_{\mu} \right)^{2} \right.\nonumber\\ 
&- &\left. iq^{2}\frac{M^4m^{2}}{\Lambda^6}\left(\partial_{\mu}\bar{\theta} + 2\pi \bar{a}_{\mu} \right)\epsilon^{\mu\nu\rho\sigma}\left( A_{\nu} + \frac{1}{q} \partial_{\nu}\theta + \frac{2\pi}{q} a_{\nu}\right)\partial_{\rho}\left( A_{\sigma} + \frac{1}{q} \partial_{\sigma}\theta + \frac{2\pi}{q} a_{\sigma}\right) +\tilde{\rho}\;cos(\bar{\theta}) \right.\nonumber\\
 &+&\left. ib_{\mu\nu}J_V^{\mu\nu} - ib_{\mu\nu}\epsilon^{\mu\nu\rho\sigma}\partial_{\rho}a_{\sigma} +i\bar{b}_{\mu\nu}\bar{J}_V^{\mu\nu} - i\bar{b}_{\mu\nu}\epsilon^{\mu\nu\rho\sigma}\partial_{\rho}\bar{a}_{\sigma}  \right) +ie\int d^4x \; A_{\mu}j_{\mu}  
\end{eqnarray} 
We note that the fundamental aspect of this result is the axion-like term. The field $\bar{a}$ in fact defines the main ingredients in the topological superconductor description. It is peculiar that the corresponding vortices, $\bar{J}_V =\varepsilon\partial \bar{a}$  don't carry electromagnetic flux, as opposed to the usual $a$ vortices. If we consider a perfect superconductor state, without vortices, $J_V = \bar{J}_V =0$, the action becomes
\begin{eqnarray}
\label{topSC2FS-c}
S_j &=& S -ie\int d^4x \; A_{\mu}j_{\mu}  \nonumber\\
&=& \int d^{4}x\; \left( \frac{1}{4}F_{\mu\nu}(A)F^{\mu\nu}(A) + \frac{q^{2}M^{2}}{2}\left( A_{\mu} + \frac{1}{q} \partial_{\mu}\theta \right)^{2} + \frac{m^{2}}{2}\left(\partial_{\mu}\bar{\theta}\right)^{2} \right.\nonumber\\ 
&- &\left. iq^{2}\frac{M^4m^{2}}{\Lambda^6} \partial_{\mu}\bar{\theta} \epsilon^{\mu\nu\rho\sigma}A_{\nu} \partial_{\rho}A_{\sigma}  +\tilde{\rho}\;cos(\bar{\theta}) \right)  +ie\int d^4x \; A_{\mu}j_{\mu}  
\end{eqnarray} 
We note that in this formulation $\theta$ and $\bar{\theta}$ are regular functions. Another way of reintroducing vortices in this expression is to take  $\theta$ and $\bar{\theta}$ as angular variables (multivalued functions). This makes contact with the formulation obtained in \cite{Qi:2012cs}, noting that $\bar{\theta} = \theta_1 -\theta_2$ and making $\frac{M^4m^{2}}{\Lambda^6}  = \frac{1}{16\pi^2}$.

Seeking for a low energy description, we integrate the massive fields in \eqref{topSC2FS-b} and retain only the lowest order terms in power of derivatives (Or, equivalently, we let $m$, $M$ $\rightarrow \infty$). The resulting effective theory is the same as the one for the usual superconductor, so that the only degrees of freedom relevant in the deep infrared seen to be the vortices and quasiparticles.    
\begin{eqnarray}
\label{topSC2FS-top2}
S_j &\rightarrow& \int d^{4}x\; \left(  ib_{\mu\nu}J_V^{\mu\nu} - ib_{\mu\nu}\epsilon^{\mu\nu\rho\sigma}\partial_{\rho}a_{\sigma}   -  i\frac{2\pi e}{q}  a_{\mu}j_{\mu} \right) 
\end{eqnarray} 
The vortices $\bar{J}_V$ have completely decoupled from the theory. But its presence can be seen as an important aspect of the formulation by noting that the supercurrent 
\begin{eqnarray}
\label{supercurrent}
j_s^{\mu} \equiv  q^2 M^2 \left( A^{\mu} + \frac{1}{q} \partial^{\mu}\theta + \frac{2\pi}{q} a^{\mu}\right)  &=& \partial_{\nu} F^{\nu\mu}  -  2iq^2\frac{M^4m^{2}}{\Lambda^6}\left(\partial_{\mu}\bar{\theta} + 2\pi \bar{a}_{\mu} \right)\epsilon^{\nu\mu\rho\sigma} \partial_{\rho}\left( A_{\sigma} + \frac{1}{q} \partial_{\sigma}\theta + \frac{2\pi}{q} a_{\sigma}\right)\nonumber\\
&+& iq^2\frac{M^4m^{2}}{\Lambda^6} 2\pi \partial_{\mu} \bar{a}_{\rho} \epsilon^{\nu\mu\rho\sigma} \left( A_{\sigma} + \frac{1}{q} \partial_{\sigma}\theta + \frac{2\pi}{q} a_{\sigma}\right)
\end{eqnarray} 
is not conserved due to the presence of the $\bar{J}_V$ vortices
\begin{eqnarray}
\label{supercurrent-anomaly}
\partial_{\mu} j_s^{\mu} =   - i2\pi q^2 \frac{M^4m^{2}}{\Lambda^6}\epsilon^{\mu\nu\rho\sigma} \partial_{\mu}\bar{a}_{\nu} \partial_{\rho}\left( A_{\sigma} + \frac{2\pi}{q} a_{\sigma}\right) = - i2\pi q^2 \frac{M^4m^{2}}{\Lambda^6} \bar{J}_V^{\rho\sigma}\partial_{\rho}\left( A_{\sigma} + \frac{2\pi}{q} a_{\sigma}\right)
\end{eqnarray} 
This non-conservation of the canonical supercurrent is the main characterization of a topological superconductor. One can note that the failure of this current conservation is localized on the $\bar{J}_V$ vortex configuration. It is interesting to note that the $\bar{J}_V$ vortices do not carry electromagnetic flux. This signals that the nature of the fermionic degrees of freedom underlying the ``anomaly'' must be Majorana modes localized on $\bar{J}_V$. A potential paradox related to the fact that uncharged degrees of freedom could not account for the inflow of charges was pointed out recently in \cite{Stone:2016pof}, where a solution was proposed arguing that the contribution from boundary terms in the computation of the equations of motion would provide the missing terms canceling the net inflow.

At this point it is important to discuss in more detail the relation between our result, represented by \eqref{topSC2FS-b}, or its lowest energy limit \eqref{topSC2FS-top2}, and the results presented in \cite{Qi:2012cs} and \cite{Stone:2016pof}, where effective actions for topological superconductors were also proposed. First of all it must be observed that in our formalism the parameters $M$, $m$ and $\Lambda$ are not microscopically determined. One of the main points in \cite{Stone:2016pof} is the determination of the coefficient of the topological term (given here by $\frac{M^4m^2}{\Lambda^6}$) that the authors point out to be $\frac 13$ of the one found in \cite{Qi:2012cs}. The difference comes from a proper consideration of the microscopic fermionic couplings and its relation to the anomaly (for details see \cite{Stone:2016pof}). In the present work we have not started from a microscopic fermionic theory and our analysis comes from studying possible vortices configurations in the system and its effects on the electromagnetic response, as dictated by the Julia-Toulouse approach. Therefore the result do not touch on the issue of the coefficient of the topological term. In that way, up to the (important) numerical factors, the action \eqref{topSC2FS-b} can be mapped to the form of the action presented in \cite{Qi:2012cs} or \cite{Stone:2016pof} adjusting $M$, $m$ and $\Lambda$ accordingly. 

On the other hand, the formulation here presented highlights and makes more precise the role of vortices in the establishment of the topological superconducting phase in the material. In fact, in \cite{Qi:2012cs} the action is written with the vortices described by multivalued fields, which makes the analysis of the effects related to the vortices configurations, such as the supposed  anomaly, less clear. In \cite{Stone:2016pof}, the authors observed that it was important to consider boundary terms related to the presence of vortices. Their strategy was to devise an effective theory for one Fermi surface only, which contains a Wess-Zumino-Witten term taking care of maintaining gauge invariance, and work out the supposed anomaly when two copies of this theory are considered, describing two Fermi surfaces of opposite Chern numbers, the same system described by \cite{Qi:2012cs}  and the present work. Their conclusion is that there is no anomaly since gauge symmetry is never broken. In order to reach this conclusion the authors argued that the anomaly would be canceled once vortices were properly taken into account by retaining boundary terms in their derivation of the current, but no explicit details of the role of the different vortices were given. Thus a difficulty in both \cite{Qi:2012cs}  and \cite{Stone:2016pof} is the explicit description of vortices in the system. In particular, the deep infrared limit \eqref{topSC2FS-top2} was not obtained in both these references and we consider it to be an important result of the present work. We see that the procedure that was undertaken here, explicitly constructing the theory through considerations of dilution and condensation of vortices configurations, allowed us to properly take into account the vortices by construction and deal with the supposed ``anomaly''. More precisely, the charge current, as defined by the functional derivative of \eqref{topSC2FS-b} with respect to the gauge field $A$, is explicitly conserved once the equation of motion of $\theta$ is taken into account (as similarly considered in \cite{Stone:2016pof}), as can be easily checked. This of course follows from the maintenance of the gauge symmetry throughout our derivation. Nevertheless, the canonical supercurrent $j_s$ defined in eq.\eqref{supercurrent}, a quantity that is normally conserved in the usual superconducting state, is here not conserved due to the presence of the $\bar{J}_V$ current. But it is important to note that this is not an anomaly in the charge current, in accordance with the results of \cite{Stone:2016pof}. As already mentioned, we consider this to be the main characterization of the topological superconductor state here discussed. Our result also establishes the important physical characterization of the system  through its lowest energy effective degrees of freedom, as encapsulated in the action \eqref{topSC2FS-top2} supplemented by the characterization of the $\bar{J}_V$ as the locus of the failure of the supercurrent conservation, associated with Majorana degrees of freedom. The clear physical description of the phases of the system also represents a fundamental aspect of our formulation. The phases are understood as functions of the $J_V$ and $\bar{J}_V$ configurations, discussed after eqs. \eqref{pathint-maxwell-6}, \eqref{action2fermi-2} and \eqref{etaaction}.

\section{The case of multiple Fermi surfaces}
The discussion of the previous section can be straightforwardly generalized to the case when the system displays $N$ Fermi surfaces. In this case, we define the total current as the normalized sum over all the Fermi surface currents
\begin{align}
  \label{currents-NFermi}
  J^{\mu} = \frac{1}{N}\sum_i^N   J_i^{\mu}
  \end{align}
The fluxes between Fermi surfaces are described by instantons $\rho_{ij}$ that work as ``holes'' allowing for electric current flow between Fermi surfaces $i$ and $j$, which can be locally described by 
\begin{align}
  \label{currents-NFermi2}
 \partial_{\mu}  J_i^{\mu} = \sum_{j\neq i} \rho_{ij}
\end{align} 
Note that the total current conservation, $\partial_{\mu} J^{\mu} = 0$, demands that $\rho_{ij} = -\rho_{ji}$. It is convenient to define the relative currents
\begin{align}
  \label{currents-NFermi3}
  \bar{J}_i^{\mu} =  J_i^{\mu} - J^{\mu}   
  \end{align}
It follows that  
\begin{align}
  \label{currents-NFermi4}
 \partial_{\mu}  \bar{J}_i^{\mu} &= \sum_{j\neq i} \rho_{ij}\nonumber\\
 \sum_i^N   \bar{J}_i^{\mu} &=0
\end{align} 
The partition function defining the system in terms of $J_{\mu}$ and $\bar{J}_{\mu}$ reads
\begin{align}
 \label{pathint-maxwell-NFS}
 Z[j] = \left(\prod_{j>i}\prod_{i}\sum_{J, \bar{J}_i, \rho_{ij}}  \int {\cal D} A\right) \delta[\partial_{\mu}J^{\mu}]\delta[\partial_{\mu}  \bar{J}_i^{\mu} -\sum_{j\neq i} \rho_{ij}]\delta[\sum_i^N   \bar{J}_i^{\mu}]e^{-\int d^4x \left( \frac{1}{4} F_{\mu\nu}F_{\mu\nu} - iqA^{\mu}J^{\mu} \right)} e^{-ie\int d^4x \; A_{\mu}j_{\mu}  }
 \end{align} 
exponentiating the constraints in the delta function through Lagrange multipliers $\theta$, $\bar{\theta}_i$ and $\lambda_{\mu}$, we obtain
\begin{align}
 \label{pathint-maxwell-NFS-2} 
 Z[j] = \left(\prod_{j>i} \prod_{i}\sum_{J, \bar{J}_i, \rho_{ij}}   \int {\cal D} A  {\cal D} \theta  {\cal D} \bar{\theta}_{i} {\cal D} \lambda \right) \; e^{-\int d^4x \left( \frac{1}{4} F_{\mu\nu}F_{\mu\nu} - iq\left( A_{\mu} + \frac{1}{q} \partial_{\mu} \theta \right)J_{\mu} - i \sum_{i} \left(\partial_{\mu}\bar{\theta}_i +\lambda_{\mu} \right)\bar{J}_{i\mu}  + i\sum_{i,j} \rho_{ij} \bar{\theta}_i\right)} e^{-ie\int d^4x \; A_{\mu}j_{\mu}  }
 \end{align} 
Noting again that only the total current couples with the gauge field, we define here the charge symmetry \eqref{currents-redef} similarly as
\begin{align}
  \label{currents-redef-NFS}
  J_{\mu} &\rightarrow  -J_{\mu}; \;\;\;
  \bar{J}_{i\mu} \rightarrow  \bar{J}_{i\mu} ;\;\;\;  \rho_{ij} \rightarrow \rho_{ij} \nonumber\\
  A_{\mu} &\rightarrow - A_{\mu} ; \;\;\;
  \theta \rightarrow - \theta ; \;\;\;
  \bar{\theta}_i \rightarrow  \bar{\theta}_i 
\end{align} 
Following the procedure as before, with the introduction of the identity \eqref{unitpoisson} for each current, defining fields $\eta$, $\bar{\eta}_i$, and through the Poisson identity \eqref{poissonformula} introducing the corresponding vortices $K_{\mu}$ and $\bar{K}_{i\mu}$. With the further use of  eq. \eqref{vortex-field}  we can represent each $K_{\mu}$ and $\bar{K}_{i\mu}$ by vortex currents $J_V$ and $\bar{J}_{Vi}$, written in terms of  $a_{\mu}$ and $\bar{a}_{i\mu}$ 
\begin{align}
 \label{pathint-maxwell-NFS-3} 
 Z[j] = \left( \prod_{j>i}\prod_{i}\sum_{J_V, \bar{J}_{iV}, \rho_{ij}}   \int {\cal D} A  {\cal D} \eta  {\cal D} \bar{\eta}_{i} {\cal D} \theta  {\cal D} \bar{\theta}_{i}  {\cal D} a {\cal D} b  {\cal D} \bar{a_i} {\cal D} \bar{b_i} {\cal D} \lambda  \right) \; e^{-S} e^{-ie\int d^4x \; A_{\mu}j_{\mu}  } e^{-S_{\eta,\bar{\eta}_i}}
\end{align} 
with
\begin{align}
 \label{action-NFS} 
 S &= \int d^4x \left( \frac{1}{4} F_{\mu\nu}(A) F_{\mu\nu}(A) - iq\left( A_{\mu} + \frac{1}{q} \partial_{\mu} \theta +\frac{2\pi}{q} a_{\mu}\right)\eta_{\mu} - i \sum_{i} \left(\partial_{\mu}\bar{\theta}_i +\lambda_{\mu}  +2\pi \bar{a}_{i\mu} \right)\bar{\eta}_{i\mu}  + i\sum_{i,j} \rho_{ij} \bar{\theta}_i\right. \nonumber\\ 
 &\left. + ib_{\mu\nu}J_V^{\mu\nu} - ib_{\mu\nu}\epsilon^{\mu\nu\rho\sigma}\partial_{\rho}a_{\sigma} +i \sum_i \bar{b}_{i\mu\nu}\bar{J}_{Vi}^{\mu\nu} - i\sum_i \bar{b}_{i\mu\nu}\epsilon^{\mu\nu\rho\sigma}\partial_{\rho}\bar{a}_{i\sigma}  \right)
 \end{align}  
Note that 
\begin{align}
 \label{instantonterm-NFS} 
 \sum_{i,j} \rho_{ij} \bar{\theta}_i =  \sum_{j>i} \sum_{i}\rho_{ij} \left( \bar{\theta}_i - \bar{\theta}_j\right)
 \end{align} 
So that Polyakov summation leads to
\begin{align}
\label{polyakov-NFS}
\left( \prod_{j>i}\prod_{i}\sum_{\rho_{ij}} \right)e^{ \int d^4x\; i\sum_{j>i} \sum_{i}\rho_{ij}(x) \left( \bar{\theta}_i(x) - \bar{\theta}_j(x)\right)}= e^{\int d^4x \; \sum_{j>i} \sum_{i}\tilde{\rho}_{ij}\;cos( \bar{\theta}_i(x) - \bar{\theta}_j(x))}.
\end{align}
Next we define $S_{\eta,\bar{\eta}_i}$ as the most general form respecting the symmetries
\begin{align}
\label{etaactionNFS}
S_{\eta,\bar{\eta}}=\frac{1}{2M^2}\eta_{\mu}\eta_{\mu}+ \sum_i\frac{1}{2m_i}\bar{\eta}_{i\mu}\bar{\eta}_{i\mu}+ \sum_i \Lambda^6_{i}\bar{\eta}_{i\mu}\epsilon^{\mu\nu\rho\sigma}\eta_{\nu}\partial_{\rho}\eta_{\sigma}
\end{align}
Integrating over  $\eta_{\mu}$ and $\bar{\eta}_{i\mu}$ we find
\begin{align}
\label{actionTSCNfermi}
S_j &=  S +ie\int d^4x \; A_{\mu}j_{\mu} \nonumber\\
&= \int d^{4}x\; \left( \frac{1}{4}F_{\mu\nu}(A)F^{\mu\nu}(A) + \frac{q^{2}M^{2}}{2}B^{2}_{\mu} + \sum_i\frac{m_i^{2}}{2}\bar{B}^{2}_{i\mu}\right. \nonumber\\
&\left. -iq^{2}M^4\sum_i \frac{m_i^{2}}{\Lambda^6_i} \bar{B}_{i\mu}\epsilon^{\mu\nu\rho\sigma}B_{\nu}\partial_{\rho}B_{\sigma} + i\sum_{j>i} \sum_{i}\tilde{\rho}_{ij}\;cos( \bar{\theta}_i(x) - \bar{\theta}_j(x))   \right. \nonumber\\ 
 &\left. + ib_{\mu\nu}J_V^{\mu\nu} - ib_{\mu\nu}\epsilon^{\mu\nu\rho\sigma}\partial_{\rho}a_{\sigma} +i \sum_i \bar{b}_{i\mu\nu}\bar{J}_{Vi}^{\mu\nu} - i\sum_i \bar{b}_{i\mu\nu}\epsilon^{\mu\nu\rho\sigma}\partial_{\rho}\bar{a}_{i\sigma}  \right) +ie\int d^4x \; A_{\mu}j_{\mu} 
\end{align} 
where the fields $B_{\mu}$ and $\bar{B}_{i\mu}$ are defined as
\begin{align}
\label{biNFS}
B_{\mu} &= A_{\mu} + \frac{1}{q} \partial_{\mu}\theta + \frac{2\pi}{q} a_{\mu}.\nonumber\\
\bar{B}_{i\mu} &= \partial_{\mu}\bar{\theta_i} +\lambda_{\mu} + 2\pi \bar{a}_{i\mu}.
\end{align}
Again we can search for the theory in the deep infrared region and we will obtain the same as \eqref{topSC2FS-top2}, with the complete decoupling of the vortices $\bar{J}_{iV}$. But also, these vortices make their appearance in the anomaly of the corresponding supercurrent
\begin{eqnarray}
\label{supercurrent-anomaly2}
\partial_{\mu} j_s^{\mu} =   - i2\pi q^2 \sum_i \frac{M^4m_i^{2}}{\Lambda_i^6}\epsilon^{\mu\nu\rho\sigma} \partial_{\mu}\bar{a}_{i\nu} \partial_{\rho}\left( A_{\sigma} + \frac{2\pi}{q} a_{\sigma}\right) = - i2\pi q^2 \sum_i \frac{M^4m_i^{2}}{\Lambda_i^6} \bar{J}_V^{i\rho\sigma}\partial_{\rho}\left( A_{\sigma} + \frac{2\pi}{q} a_{\sigma}\right)
\end{eqnarray} 
where use has been made of the fact that $\sum_i \frac{m_i^{2}}{\Lambda_i^6} = 0$, that follows from \eqref{currents-NFermi4}.

The same analysis about the dilution and condensation of the currents discussed in the case of two Fermi surfaces stands in this case. Note that it is necessary that only one of the currents $\bar{J}_{iV}$ be nonzero in order for the anomaly to occur.

% % % % % % % % % % % % % % % % % % % % % % % % % % % % % % % % %

\section{Conclusions}
In this work we have analyzed superconducting systems with multiple Fermi surfaces. Our main purpose was to obtain effective low energy field theories describing the relevant excitations of these systems. The approach we took, inspired by the analyses of \cite{Hansson:2004wca}, led us to the result first obtained in \cite{Qi:2012cs}, where an effective axionic electromagnetic theory was proposed as the description a time reversal invariant topological superconductor. This connection amplifies the results of \cite{Hansson:2004wca} helping to solidify the idea that a superconductor state is better described as a topologically ordered state, instead of usual Ginzburg-Landau symmetry breaking characterization by an order parameter. Furthermore, we have shown that one of the main ingredients characterizing the system with multiple Fermi surfaces is the presence of vortices configuration that do not carry electromagnetic flux. These vortices decouple at the level of the deep low energy action of the system, in the sense that the effective theory for the electromagnetic response of the system is the same as for one Fermi surface only. Nevertheless the vortices contribute non-trivially triggering the non-conservation of the canonical supercurrent, since the source of this non-conservation resides in a vortex without flux, one can view it as induced by Majorana modes localized on the vortex. This provides an important characterization of this particular superconducting state.

\section*{Acknowledgments}

The Conselho Nacional de Desenvolvimento Cient\' ifico e Tecnol\'ogico (CNPq-Brazil), the Coordena\c{c}\~ao de Aperfei\c{c}oamento de Pessoal de N\' ivel Superior (CAPES) and Funda\c{c}\~ao de Amparo a Pesquisa do Rio de Janeiro (FAPERJ) are acknowledged for financial support. M. S. Guimaraes thanks SR2-UERJ, CNPq and FAPERJ for financial support. \footnote{M.S. Guimaraes is supported by the Jovem Cientista do Nosso Estado program - FAPERJ E-26/202.844/2015, is a level PQ-2 researcher under the program Produtividade em Pesquisa - CNPq, 307905/2014-4 and is a Procientista under SR2-UERJ.} D.R. Granado is grateful for a PDSE scholarship from CAPES.

\appendix 
\section{Proof of the relation eq.\eqref{vortex-field}}
\label{proof}
%%%%%%%%%%%%%%%%%%%%%%%%%%%
\noindent For a $p$-form $a^{\mu_1\dots\mu_p}$ that satisfies the gauge fixing condition $\partial_{\mu_1}a^{\mu_1\dots\mu_p}=0$, consider the equation:
\begin{equation}
\epsilon^{\mu_1\dots\mu_{p}\mu_{p+1}\mu_{p+2}\dots\mu_d}\partial_{\mu_{p+1}}a_{\mu_1\dots\mu_{p}}=J^{\mu_{p+2}\dots\mu_d}
\end{equation}
From this we uniquely get:
\begin{equation}
a_{\mu_1\dots\mu_p}=\frac{1}{p!~c_{d-p-1}}\frac{1}{\partial^2}\epsilon_{\mu_1\dots\mu_p\mu_{p+1}\mu_{p+2}\dots\mu_d}\partial^{\mu_{p+1}}J^{\mu_{p+2}\dots\mu_d}
\end{equation}
where $c_{d-p-1}$ is a constant. Therefore we have:
\begin{eqnarray}
&&\sum_{J_{v}}\int\mathcal{D}a_{GF}\delta( J^{\mu_{p+2}\dots\mu_d} -\epsilon^{\mu_1\dots\mu_{p}\mu_{p+1}\mu_{p+2}\dots\mu_d}\partial_{\mu_{p+1}}a_{\mu_1\dots\mu_{p}} )f(a)\nonumber\\
&=&\sum_{J_{v}}\int\mathcal{D}a_{GF}\delta\left(a_{\mu_1\dots\mu_p}-\frac{1}{p!~a_{d-p-1}}\frac{1}{\partial^2}\epsilon_{\mu_1\dots\mu_p\mu_{p+1}\mu_{p+2}\dots\mu_d}\partial^{\mu_{p+1}}J^{\mu_{p+2}\dots\mu_d}\right)f(a)
\end{eqnarray}
where $a_{GF}$ stands for the gauge fixed field $a$. As we are considering the integration measure to be gauge fixed,  we do not need to worry about zero modes in the Jacobian (which is the determinant of the operator $\epsilon\partial$). Therefore:
\begin{eqnarray}
&&\sum_{J_{v}}\int\mathcal{D}a_{GF}\delta( J^{\mu_{p+2}\dots\mu_d} -\epsilon^{\mu_1\dots\mu_{p}\mu_{p+1}\mu_{p+2}\dots\mu_d}\partial_{\mu_{p+1}}a_{\mu_1\dots\mu_{p}} )f(a)\nonumber\\
&\sim&\sum_{J_{v}}f\left(\frac{1}{\partial^2}\epsilon_{\mu_1\dots\mu_p\mu_{p+1}\mu_{p+2}\dots\mu_d}\partial^{\mu_{p+1}}J^{\mu_{p+2}\dots\mu_d}\right)\nonumber\\
&\sim&\sum_Kf(K)
\end{eqnarray}
where the last step follows from the equation \eqref{vc}. It is understood that the sum over $K$ spans only the configurations with $\partial K=0$, so that $K\propto\frac{1}{\partial^2}\epsilon\partial J$.

\end{document}